\RequirePackage{ifpdf}
\ifpdf 
\documentclass[pdftex]{sigma}
\else
\documentclass{sigma}
\fi

\begin{document}


\renewcommand{\PaperNumber}{019}

\FirstPageHeading

\renewcommand{\thefootnote}{$\star$}

\ShortArticleName{Single-Pole Interaction of the Particle with the String}

\ArticleName{Single-Pole Interaction of the Particle with the String\footnote{This
paper is a contribution to the Proceedings of the Seventh
International Conference ``Symmetry in Nonlinear Mathematical
Physics'' (June 24--30, 2007, Kyiv, Ukraine). The full collection
is available at
\href{http://www.emis.de/journals/SIGMA/symmetry2007.html}{http://www.emis.de/journals/SIGMA/symmetry2007.html}}}

\Author{Milovan VASILI\'C and Marko VOJINOVI\'C}

\AuthorNameForHeading{M. Vasili\'c and M. Vojinovi\'c}

\Address{Institute of Physics, P.O.Box 57, 11001 Belgrade, Serbia}
\Email{\href{mailto:mvasilic@phy.bg.ac.yu}{mvasilic@phy.bg.ac.yu}, \href{mailto:vmarko@phy.bg.ac.yu}{vmarko@phy.bg.ac.yu}}

\ArticleDates{Received October 24, 2007, in f\/inal form January 20, 2008; Published online February 12, 2008}

\Abstract{Within the framework of generalized Papapetrou method, we derive the ef\/fective
equations of motion for a string with two particles attached to its ends,
along with appropriate boundary conditions. The equations of motion are the
usual Nambu--Goto-like equations, while boundary conditions turn out to be
equations of motion for the particles at the string ends. Various properties
of those equations are discussed, and a simple example is treated in detail,
exhibiting the properties of Neumann and Dirichlet boundary conditions and
giving a small correction term to the law of Regge trajectories due to the
nonzero particle mass.}


\Classification{83C10; 83C55; 81T30; 70S10}

\section{Introduction}

The interest in studying extended objects in high energy
physics began with the observation that meson resonances could be viewed as
rotating relativistic strings \cite{Nambu,Goto}. This model provided a
successful explanation of Regge trajectories. Nevertheless, the model assumes
that the two quarks attached to the string have zero mass and zero angular
momentum. The purpose of this short note is to take into account small mass
of the two particles, and exhibit some basic properties of equations of motion
in this case.

In order to do so, one needs to derive the equations of motion for the string
with particles attached to its ends. The general method that lends itself for
doing this was developed in~\cite{Vasilic2006}, and represents the
generalization of the Mathisson--Papapetrou method
\cite{Mathisson1937, Papapetrou1951} to include extended objects.

The layout of the paper is the following. In Section~\ref{sec2} we formulate
the stress-energy tensor for the string with two particles attached in the
simplest, single-pole approximation. Then we utilize its covariant
conservation
to derive equations of motion and boundary conditions for the string. It turns
out that the boundary conditions are nothing but equations of motion for the
particle residing at the end of the string. We recognize the force that acts
upon the particle, and comment that in case of Nambu--Goto string this force
has
the form similar to Lorentz force from electrodynamics. We also comment on the
particle mass conservation and the possibility of writing an action for the
system of equations of motion.

In Section~\ref{sec3} we proceed to f\/ind an example solution of the equations
of motion, where the string is of Nambu--Goto type and lying along a straight
line of constant length $L$ while rotating with constant angular velocity
about
its center. The velocity of the particles at the ends turns out to depend on
the length of the string, the mass of the particle and the string tension, in
such a way that it is always less than the velocity of light, as expected. We
discuss two limiting cases~-- in the limit of zero particle mass, the
particle
velocity becomes equal to the speed of light, and represents the situation
analogous to imposing Neumann boundary conditions in usual string theory. In
the
limit of inf\/inite particle mass, its velocity equals zero, and represents the
situation analogous to imposing Dirichlet boundary conditions where the string
is attached to a $0$-brane, i.e.\ a particle. Other cases, those with f\/inite
particle velocity, do not appear in usual string theory.

Section~\ref{sec4} is devoted to computing the total energy and angular
momentum
for the example system discussed above. By eliminating the string length
parameter
from the equations, in the limit of small particle masses one recovers the
familiar law of Regge trajectories, but now with a small correction term due
to masses being small but still nonzero. This result is a demonstration
how one can calculate corrections to the Regge law due to the presence of
particles at the string ends.

In Section \ref{sec5} we give our f\/inal comments and conclusions.

Our conventions are the same as in \cite{Vasilic2006}, aside from the metric
signature, which we now take to be ${\rm diag \,}(-,+,+,+)$.
Greek indices $\mu,\nu,\dots$ are the spacetime
indices, and run over $0,1,\dots,D-1$. Latin indices $a,b,\dots$ are the
world
sheet indices and run over $0,1,\dots,p$. The Latin indices $i,j,\dots$
refer
to the world sheet boundary and take values $0,1,\dots,p-1$. The coordinates
of spacetime, world sheet and world sheet boundary are denoted by~$x^{\mu}$,
$\xi^a$ and~$\lambda^i$, respectively. The corresponding metric tensors
which
are used to lower indices are denoted by~$g_{\mu\nu}(x)$,
$\gamma_{ab}(\xi)$
and $h_{ij}(\lambda)$, while the indices are raised by the inverse metrics
$g^{\mu\nu}$, $\gamma^{ab}$ and $h^{ij}$. We shall mainly be interested
in
the realistic case $D=4$.

\section{Equations of motion}\label{sec2}

We begin by considering an isolated system consisted of a
string with two massive particles attached to its ends, in Riemannian
$4$-dimensional spacetime. The basic starting point of the derivation of
equations of motion is the covariant conservation of the symmetric
stress-energy tensor \cite{Papapetrou1951,Vasilic2006}:
\[
\nabla_{\nu}T^{\mu\nu}=0  .
\]
The stress-energy tensor is written as a sum of two terms,
\[
T^{\mu\nu} = \int_{{\cal M}} d^2\xi \sqrt{-\gamma}
B_{\rm s}^{\mu\nu} \frac{\delta^{(4)}(x-z)}{\sqrt{-g}} +
\int_{\partial{\cal M}} d\lambda \sqrt{-h}
B_{\rm p}^{\mu\nu}\frac{\delta^{(4)}(x-z)}{\sqrt{-g}} ,
\]
representing the stress-energy components for the string and
the particles at string endpoints, respectively. We restrict ourselves to the
simplest, single-pole approximation, neglecting any higher terms containing
derivatives of the Dirac delta function. The procedure described in
\cite{Vasilic2006}
then yields the familiar world sheet equations
\begin{gather} \label{jna1}
\nabla_a \big( m^{ab} u_b^{\mu} \big) =0  ,
\end{gather}
as well as expressions that determine the form of the stress-energy tensor:
\begin{gather} \label{jna2}
B_{\rm s}^{\mu\nu} = m^{ab} u_a^{\mu} u_b^{\nu}, \qquad
B_{\rm p}^{\mu\nu} = m v^{\mu} v^{\nu}.
\end{gather}
Here the world sheet is described by the parametrized equations
$x^{\mu} = z^{\mu}(\xi^a)$, where $a\in\{ 0,1 \}$. Similarly, the world
sheet
boundary is described by the parametrized equations $\xi^a =
\zeta^a(\lambda)$
or equivalently $x^{\mu}=z^{\mu}(\zeta^a(\lambda))$. Consequently, we def\/ine
the notation for the vectors tangent to the world sheet and a vector tangent
to the boundary:
\[
u_a^{\mu} \equiv \frac{\partial z^{\mu}}{\partial \xi^a}, \qquad
v^a \equiv \frac{\partial \zeta^a}{\partial \lambda}, \qquad v^{\mu} =
u_a^{\mu}v^a.
\]
Aside from equations of motion, one also gets the appropriate boundary
conditions
for the string. The particle part of $T^{\mu\nu}$ is constrained by the
requirement
that particle trajectories coincide with the string boundary, so the resulting
boundary conditions are:
\begin{gather} \label{jna3}
\frac{D}{ds} \left( mv^{\mu} \right) = n_a m^{ab} u_b^{\mu}  .
\end{gather}
These boundary conditions are written using the parametrization $\lambda=s$
(where $s$ is the proper distance parameter), or equivalently f\/ixing the
gauge
condition $h\equiv v_{\mu}v^{\mu}=-1$, and are reinterpreted as the particle
equations of motion. The stress-energy tensor of such a system has also been
studied in \cite{Carter1990}.

The boundary conditions represent the equation determining particle
trajectory,
and the term on the right-hand side represents the force that string exerts on
the particle. If the boundary~$\partial{\cal M}$ consists of two disjoint
lines,
$\partial{\cal M} = \partial{\cal M}_1 \cup \partial{\cal M}_2$ and
$\partial{\cal M}_1 \cap \partial{\cal M}_2 = \varnothing$,
the masses $m$ of the two particles may dif\/fer.

In general, equations \eqref{jna1} and \eqref{jna3} cannot be derived by
extremizing some action if one does not introduce auxiliary variables in
the theory. However, for the case of Nambu--Goto matter described by the
choice
\[
m^{ab} = T \gamma^{ab}  ,
\]
such an action exists, and is of the form
\[
S = T \int_{{\cal M}} d^2\xi\,   \sqrt{-\gamma} - m \int_{\partial{\cal M}}
d\lambda \,  \sqrt{-h}   .
\]
The information that the particles are attached to string ends is encoded in
the
requirement that the domain of integration for the particle action be
precisely
the boundary of the string world sheet ${\cal M}$. As above, if the boundary
consists
of two disjoint lines, the particle part of the action can be
split in two independent parts, each containing independent mass parameter,
$m_1$ and~$m_2$, which need not be equal. This action can be also used to
model monopole-antimonopole pairs connected by Nambu--Goto strings, formed in
phase transition in the early universe \cite{Berezinsky1997, Martin1997}.

There are two other peculiarities of the Nambu--Goto choice of matter, as we
shall show.
First, by contracting the equation \eqref{jna3} with $v_{\mu}$, and having
in mind
the choice $m^{ab} = T\gamma^{ab}$, one easily derives the law of mass
conservation:
\[
\frac{dm}{ds} =0.
\]
In the general case of $m^{ab}$ this need not hold, because matter may be
allowed to
f\/low from the particle to the interior of the string and vice versa.

Second, denoting the right-hand side of \eqref{jna3} as the force $F^{\mu}$,
we can rewrite it as:
\[
F^{\mu} \equiv T n^{\mu} = T F^{\mu}{}_{\nu}v^{\nu},
\qquad F^{\mu\nu} \equiv u_a^{\mu}e^{ab}u_b^{\nu}.
\]
Here $e^{ab}$ is the covariant totally antisymmetric Levi-Civita tensor. In
this
notation, the equation of motion for the particle takes the form:
\[
m\frac{D}{ds}v^{\mu} = T F^{\mu\nu}v_{\nu},
\]
which looks identically the same as the well known Lorentz force law from
ordinary
electrodynamics:
\[
m\frac{D}{ds}v^{\mu} = q {\cal F}^{\mu\nu}v_{\nu},
\]
${\cal F}^{\mu\nu}$ being the antisymmetric electromagnetic f\/ield strength
tensor, and $q$ the appropriate charge of the particle. While the
``string f\/ield strength tensor'' $F^{\mu\nu}$ is also antisymmetric, it is of
course
of entirely dif\/ferent nature, but the identical form of the force law
nevertheless
seems interesting.

\section{Neumann and Dirichlet boundary conditions}\label{sec3}

In what follows, we shall assume that the string is made of the Nambu--Goto
type of matter, moving in a $4$-dimensional f\/lat spacetime:
\[
m^{ab} = T \gamma^{ab} , \qquad {R^{\mu}}_{\nu\lambda\rho}=0  .
\]
Then, the world sheet equations \eqref{jna1} reduce to the familiar
Nambu--Goto equations, and the third term on the right-hand side of~\eqref{jna3} becomes~$T n^{\mu}$.

Now, we look for a simple, \emph{straight line} solution of the equations of
motion \eqref{jna1}. This has also been discussed in~\cite{Martin1997}, but
for dif\/ferent purpose. Without loss of generality, we put
\[
\vec{z} = \vec{\alpha} (\tau) \sigma ,\qquad z^0=\tau  ,
\]
where $\xi^0\equiv \tau$ and $\xi^1\equiv \sigma$ take values in the
intervals
$(-\infty,\infty)$ and $[-1,1]$, respectively. This choice represents a
string
lying along the line of length $L$, in appropriate coordinates, with its
center
point at rest in coordinate origin.
 Assuming that the string length
$L=2|\vec{\alpha}|$, and the velocity of the string ends
$V=|d\vec{\alpha}/d\tau|$ are constant, the equation~\eqref{jna1}
reduces to
\[
\frac{d^2}{d\tau^2}\vec{\alpha} + \omega^2\vec{\alpha}=0 , \qquad
\omega\equiv \frac{2V}{L} .
\]
It describes uniform rotation in a plane. Choosing the rotation plane to be
the $x-y$ plane, we get the solution
\[
\vec{\alpha} = \frac{L}{2} \left( \cos \omega\tau  \vec{e}_x +
\sin \omega\tau \,\vec{e}_y \right) .
\]

Next we consider the boundary equation~\eqref{jna3}. Omitting the details of
the calculation, we f\/ind that the particle velocity becomes
\begin{gather} \label{jna4}
V = \frac{1}{\sqrt{1+\frac{2m}{TL}}}  .
\end{gather}
Of course, our assumption that the velocities of two ends are equal implies
that in this case masses of two particles must be equal.

By inspecting the expression \eqref{jna4}, we see that $V<1$, as it should
be. In the limit $m\to 0$, the string ends move with the speed of light,
representing the Nambu--Goto dynamics with Neumann boundary conditions, i.e.\ a
free string. When $m\to\infty$, the string ends do not move. This is an
example
of Dirichlet boundary conditions, where string ends are attached to a f\/ixed
$p$-brane, in our case $p$ being equal to zero, i.e.\ the particle. All
other
choices for $m$ represent cases outside these two, and do not appear in
ordinary
string theory.

\section{Regge trajectories law}\label{sec4}

The total angular momentum and energy of the example system
considered in previous section are calculated using the usual def\/initions:
\[
E = \int d^3 x \, T^{00}, \qquad
J^{\mu\nu} = \int d^3x  \, x^{[\mu}T^{\nu]0}.
\]
Substituting the appropriate solution into the expressions \eqref{jna2} and
back
into the stress-energy tensor, after integration one easily f\/inds
\begin{gather}
E=TL\frac{\arcsin V}{V} + \frac{2m}{\sqrt{1-V^2}} ,
\qquad
J = \frac{TL^2}{4} \left( \frac{\arcsin V}{V^2} - \frac{\sqrt{1-V^2}}{V}
\right) + \frac{2m}{\sqrt{1-V^2}} \frac{LV}{2}   .
\nonumber
\end{gather}
These equations have obvious interpretation. The total energy of the system
consists of the string energy and the kinetic energy of the two particles,
while
the total angular momentum is the sum of the string orbital angular momentum
and
orbital angular momenta of the two particles.

In the limit of small particle masses, the free parameter $L$ can be
eliminated in favour of $E$, which leads to
\begin{gather*} 
J = \frac{E^2}{2\pi T} - \frac{4}{3T} \sqrt{\frac{E}{\pi}} m^{\frac{3}{2}} +
{\cal O}(m^2) .
\end{gather*}
The f\/irst term on the right-hand side def\/ines the known Regge trajectory,
while the second represents a small correction due to the presence of
massive particles at the string ends. As we can see, the unique Regge
trajectory of the ordinary string theory splits into a family of
distinctive trajectories, and that $J$ becomes nonlinear in $E^2$.

Of course, this result is specif\/ic to the particular example discussed above,
but it
shows that one can in principle calculate corrections to the law of Regge
trajectories
by allowing the end-point particles to have nonzero mass.

\section{\label{sec5}Concluding remarks}

In this paper we have analyzed the system consisting of a string
with two particles attached to its ends. The method we use is a generalization
of the
Mathisson--Papapetrou method for pointlike matter
\cite{Mathisson1937,Papapetrou1951}. It has already been used in
\cite{Vasilic2006} for the derivation of equations of motion of
extended objects. Using those results, we have derived the equations of motion
for the string along with the appropriate boundary conditions. These boundary
conditions turn out to be the equations of motion for the two particles
attached
to the string ends.

These equations of motion display the string force acting on the particle. In
some cases
they imply that the mass of the particle is conserved, and the force term can
be rewritten
in the form that is formally identical to Lorentz force law of
electrodynamics. Also, in
general case the equations of motion for the string and the particle do not
allow themselves
to be derived by extremizing some action, without introduction of auxiliary
variables.
However, in the special case of Nambu--Goto matter for the string, such an
action does exist.

Next we specialized to the case of the usual Nambu--Goto string with two
massive
particles at its ends. The equations of motion can be solved exactly for the
case of a
straight line string rotating around its center. It turns out that the
velocity of the
string ends is less than the velocity of light, and is dependent on the masses
of the particles. In this way, one is provided a~way to describe both Neumann
and
Dirichlet boundary conditions for the Nambu--Goto string in the limits where
the masses of
the two particles approach zero or inf\/inity, respectively.

Finally, given this solution, one can calculate the total energy and angular
momentum of
the system, and in the limit of small particle masses derive the relation
connecting the
total angular momentum with the total energy of the string and the particles.
This relation
represents the law of Regge trajectories, with a correction term due to the
particle masses.
In this setting, there is not only one Regge trajectory, but a whole family,
due to
arbitrariness in choice of mass parameters for the constituent particles.
Also, the Regge
trajectory ceases to be linear in $E^2$ in the limit of small but nonzero
particle masses.

We remark at the end that more general conf\/igurations can also be treated using
this forma\-lism.
It is an open question, however, whether the equations of motion of those
conf\/igurations
are integrable for some special case of motion. Also, the whole treatment presented
in this paper is entirely classical, and one could in principle study the
quantum theory of strings with particles attached to its ends. The spectra of
perturbations of such strings is also an open question. Both these questions
represent possible topics for future research.

\subsection*{Acknowledgments}

This work was supported by the Serbian Science Foundation, Serbia.
A CEI grant for participation in the Seventh International Conference
``Symmetry in Nonlinear Mathematical Physics'' is  gratefully acknowledged.

\pdfbookmark[1]{References}{ref}
\LastPageEnding

\end{document}